\documentclass[twocolumn,12pt]{article}

\usepackage[authoryear]{natbib}
\usepackage{amssymb}
\usepackage{graphicx}
\usepackage{epstopdf}

\begin{document}

\title{Automatic quantitative morphological analysis of interacting galaxies}

\author{Lior Shamir\footnote{lshamir@mtu.edu} \\ Lawrence Technological University, Southfield, Michigan, USA \\ 
Anthony Holincheck \\ George Mason University, Fairfax, Virginia, USA \\
John Wallin \\ Middle Tennessee State University, Murfreesboro, Tennessee, USA
}

\maketitle

\begin{abstract}
The large number of galaxies imaged by digital sky surveys reinforces the need for computational methods for analyzing galaxy morphology. While the morphology of most galaxies can be associated with a stage on the Hubble sequence, morphology of galaxy mergers is far more complex due to the combination of two or more galaxies with different morphologies and the interaction between them. Here we propose a computational method based on unsupervised machine learning that can quantitatively analyze morphologies of galaxy mergers and associate galaxies by their morphology. The method works by first generating multiple synthetic galaxy models for each galaxy merger, and then extracting a large set of numerical image content descriptors for each galaxy model. These numbers are weighted using Fisher discriminant scores, and then the similarities between the galaxy mergers are deduced using a variation of Weighted Nearest Neighbor analysis such that the Fisher scores are used as weights. The similarities between the galaxy mergers are visualized using phylogenies to provide a graph that reflects the morphological similarities between the different galaxy mergers, and thus quantitatively profile the morphology of galaxy mergers.
\end{abstract}

{\bf keywords}:
galaxies: structure -- galaxies: evolution -- methods: analytical -- techniques: image processing


\section{Introduction}
Galaxy mergers are linked to multiple forms of galactic activities such as star formation \citep{Mat07,Bri07}, quasars \citep{Hop05}, active galactic nuclei \citep{Spr08}, and galaxy morphology \citep{Spr05,Bow06}. While most single galaxies can be associated with a stage on the Hubble sequence, the morphology of a pair of interacting galaxies is more complex than the morphology of a single galaxy, making morphological analysis and classification of galaxy mergers a challenging task that requires compound catalogues and classification schemes \citep{Arp66,Str99,Sch90}. The morphology of galaxies in these catalogs is determined by manual observation, and can therefore be ambiguous. Classes defined by \cite{Arp66} and by \cite{VV1, VV2} differ in their characterization of interacting pairs. In some cases, even the class definitions can be ambiguous. For example the VV classes of “pair of coalescents’’ and “pair in contact’’ are difficult to distinguish. Even with simple galaxy morphology classifications, Galaxy Zoo has shown that some systems are intrinsically harder to classify into simple spiral and elliptical categories \citep{GZ}.

Until now, automatic tools for galaxy morphological analysis have focused on single, non-interacting, galaxies. Current approaches include parametric model-driven methods such as GALFIT \citep{Pen02}, GIM2D \citep{Sim98}, and Ganalyzer \citep{Sha11a,Sha11b}, as well as machine learning methods \citep{Bal04,Bal08,Sha09b,Ban10,Hue11}. However, these methods are based on supervised machine learning, which automates a human guided classification of objects into one of several pre-defined distinct classes. GALFIT has been used with some success to model irregular and interacting galaxies \citep{Pen10} but only by fitting over 100 parameters in the best-fit models. Handcrafting these models for large numbers of interacting galaxies and those with irregular morphologies remains a daunting task. Other proposed methods include CAS \citep{Con03}, and the Gini coefficient method \citep{Abr03} that was also applied to galaxy images to deduce the statistics of broad morphology of galaxy mergers \citep{Lot08}. 

Unlike supervised machine learning, unsupervised machine learning is not based on existing knowledge and pre-defined training data, but aims at analyzing given data to automatically deduce its properties and structural descriptors. That is, in unsupervised learning the data are processed with no prior assumptions or human guidance to detect subsets of samples that are similar to each other, outliers, etc’. In this paper we describe a method that can profile the morphology of interacting galaxies, and automatically deduce the similarities between galaxy mergers based on the galaxy images.
Simulation remains an important tool for studying the morphology of interacting galaxies. The Zooniverse project sponsored the Merger Zoo to study dozens of interacting galaxies by having Citizen Scientist volunteers attempt to simulate specific pairs of galaxies. They were tasked with identifying and evaluating the results of galaxy simulations based on how well they matched the target images. The volunteers selected over 50,000 simulations during the course of the project. These simulated images form the population of training data used in this study.

\section{Data}
\label{data}

The data used in the experiment are 54 images of interacting galaxies taken by Sloan Digital Sky Survey \citep{Yor00}. Additionally, the 50,000 simulated images of these galaxies produced by the Merger Zoo projected are sampled for training data. The images were of size 512$\times$512 pixels, downscaled to 256$\times$256 to reduce the response time of the computation process.

\section{Generating simulations of interacting galaxies}
\label{simulation}

To effectively profile the morphology of a galaxy merger, there is a need for multiple images of each galaxy so that the pattern of morphological features can be deduced. For that purpose, multiple simulated galaxy mergers were generated for each of the galaxy images using a restricted three-body simulation code called SPAM. The restricted three-body approach uses a static galaxy potential and massless ``test particles’’ to produce the tidal features seen in galaxy interactions. These models do not reproduce the gas dynamics or star formation associated with real interactions. However, they have been useful in modeling the orbital interactions and basic morphology of interacting systems. The computational speed of the code allows thousands of runs to be completed in the time required for a moderate resolution run using a treecode such as Gadget \citep{Spr00}.

The initial conditions needed to simulate the gravitational forces in an encounter between two interacting galaxies cannot be fully determined from observational data alone. A multi-year Citizen Science project called Galaxy Zoo - Mergers was undertaken to enlist the help of volunteers in determining simulation initial conditions for the 54 SDSS galaxy mergers studied here. The SPAM code would run on sets of randomly selected initial conditions. The volunteers would indicate which simulations were a possible morphological match to the target galaxies.

Subsequent rounds of review and evaluation would assign a fitness score to each simulation based upon how well it matched the target image morphology. A perfect fit would be assigned a fitness of 1, and poor fits would have a fitness as low as 0. The sets of simulations used here all had a fitness of $\sim$0.21 or greater. Details of this project are contained in \citep{Hol13}. 

For this study, a set of simulated images were generated using 20,000 massless test particles.The particles were assigned to each galaxy in proportion to the specified masses. The final positions of the test particles were used to build up a grayscale image. For each particle, a Gaussian kernel was computed and the intensity of the pixel was set based upon the value of the kernel. Multiple particle activations were added together. The activation values for all of the pixels were then adjusted so that there were a total of 255 steps on a logarithmic scale between the lowest activation and the highest activation.

The use of simulated images to capture the morphological features of interacting systems is somewhat unusual. However, the Merger Zoo project has provided an opportunity to create a set of systems that share characteristics with the original galaxy. A more standard technique of scaling, rotating, and smoothing galaxy images would provide an alternative way of obtaining these data. However, the use of these models effectively allows the inclusion of the collisional history of these systems to be part of the classification process.

The output images are in the lossless TIFF format, and each image is rotated by a random number of degrees so that the machine learning will not be biased by the absolute positions of the two interacting galaxies, which are constant across all simulated galaxy mergers generated by SPAM for a certain image of interacting galaxies.
Source code of the SPAM galaxy merger simulator is available for free download from the Astrophysics Source Code Library at http://coms.cs.mtsu.edu/jspam.

\section{Unsupervised learning of galaxy morphology}
\label{pattern_recognition}

Interacting galaxies feature complex morphology, and therefore comprehensive morphological analysis of galaxy mergers should be based on multiple numerical image content descriptors that reflect the image content. The feature set used in the analysis is the {\it WND-CHARM} scheme \citep{Sha08a,Sha08b}, which is based on extracting a very large number of image features, and was originally designed for automatic morphological analysis of cell and tissue images \citep{Sha08b,Sha09a}. The comprehensiveness of the feature set and its ability to measure very many different aspects of the visual content allows it to analyze complex morphology such as visual art \citep{Sha10,Sha12a}, and was also shown to be effective in the analysis of galaxy images \citep{Sha09b,Sha12b}.

{\it WND-CHARM} first extracts from each image a vector of 1025 numerical image content descriptors that include high-contrast features (object statistics, edge statistics, Gabor filters), textures (Haralick, Tamura), statistical distribution of the pixel values (multi-scale histograms, first four moments), factors from polynomial decomposition of the image (Chebyshev statistics, Chebyshev-Fourier statistics, Zernike polynomials), Radon features and fractal features. These features are computed from the raw pixels, but also from image transforms as well as multi-order transforms. These transform include the Fourier transform, Chebyshev transform, and Wavelet transform, as well as tandem combinations of these transforms. A detailed description of these image content descriptors and the image transforms is available in \citep{Sha08,Sha08a,Sha08b,Sha09a,Sha10,Sha12b}.

After the image features are computed, the simulated galaxy images of each target image are separated randomly into training and test sets such that 210 images are allocated for training and 20 for testing, and each of the 1025 features computed on the training set is assigned a Fisher discriminant score \citep{Bis06}. Since not all image content descriptors are expected to be informative, the features are ordered by their Fisher discriminant score, and 85\% of the features with the lowest scores are rejected in order to filter non-informative image features. 

The reason for computing the entire feature set before rejecting most features is that {\it WND-CHARM} is a data-driven algorithm, and therefore the informativeness of the features is determined statistically based on the data being processed. That is, {\it WND-CHARM} does not know which features are more informative before computing them, and therefore needs to compute all feature values for each image so that the most informative features can be selected. That approach of using a comprehensive set of numerical image content descriptors that reflect very many aspects of the visual content and then statistically selecting the most relevant features allows using the system without making any prior assumptions about the physical characteristics of the galaxies.

As described in \citep{Sha08a,Orl08}, the features are computed in groups, so that if a certain feature is needed the entire group needs to be computed, and there is no straightforward way to compute the feature without computing its group. Therefore, computing just the features used in the analysis will not lead to any improvement in the response time of the system, unless using a very small subset of the features ($\sim$5\% or less). In any case, when the ‘’classify” method of {\it WND-CHARM} is used for classifying a new galaxy just the required features are computed \citep{Sha08a}, which will lead to noticeably improved response time when not many features are used.

The similarity between each pair of two images can be estimated by the weighted distance between two image feature vectors {\it X} and {\it Y} as described by Equation~\ref{wnn}
\begin{equation}
d=\sqrt{\sum_{f=1}^{|X|} W_f(X_f-Y_f)^2},
\label{wnn}
\end{equation}
where $W_f$ is the assigned Fisher score of feature {\it f}, and {\it d} is the computed weighted distance between the two feature vectors. The predicted class of a given test image is determined by the class of the training image that has the shortest weighted distance {\it d} to the test image.

The purpose of the algorithm is not to classify the simulated galaxy mergers, but to use the simulated images to determine the similarities between the target interacting galaxies that were used to generate them. The similarity between a test simulated image and a target image of interacting galaxies is determined by first computing a vector of size {\it N} (N is the total number of target galaxy merger images), such that each entry {\it c} in the vector represents the computed similarity of the feature vector to the class {\it c}, deduced using Equation~\ref{marginal_probs},
\begin{equation}
\label{marginal_probs}
M_{f,c} = \frac{1}{\min(D_{f,c}) \cdot \sum_{i=1}^{N} \frac{1}{\min(D_{f,i})}}
\end{equation}
where $M_{f,c}$ is the computed similarity of the simulated merger {\it f} to the target merger {\it c}, and $\min(D_{f,c})$ is the shortest weighted Euclidean distance between the feature vector {\it f}, computed using Equation~\ref{wnn}.
Averaging the similarity vectors of all simulated merger images generated for a certain target galaxy merger image provides the similarities between that target merger image and any of the other target images of interacting galaxies. Repeating this for all target merger images results in a similarity matrix that represents the similarities between all pairs of target merger images. The similarity matrix contains two similarity values for each pair of target merger images. I.e., the cell $n,m$ is the similarity value between class {\it n} to class {\it m}, which may be different from the cell $m,n$. Although these two values are expected to be close, they are not expected to be fully identical due to the different images used when comparing {\it n} to {\it m} and {\it m} to {\it n}. Averaging the two values provides a single distance between each pair of target merger images \citep{Sha10}.

The distances are then visualized by using phylogenies inferred automatically by the Phylib package \citep{phylip}, which visualizes the morphological similarities between the target images of interacting galaxies. Source code of the {\it WND-CHARM} algorithm is available for free download at \newline http://vfacstaff.ltu.edu/lshamir/downloads/ImageClassifier.
While Wndchrm has the ability to reflect very complex image morphology \citep{Sha10, Sha12a}, unsupervised analysis of galaxy morphology, and especially galaxy mergers, can be a challenging problem for computing machines. Therefore, Wndchrm might not be effective in the analysis of some morphological types. For instance, Wndchrm was found ineffective in determining the rotation directionality of spiral galaxies, and therefore model-driven tools were developed and used for that task \citep{Sha11a}. Also, Wndchrm is not scale invariant, and all images processed by Wndchrm should be of the same size. In any case, when using Wndchrm the user should use the classification accuracy as an indication of the informativeness of Wndchrm in the context of the image analysis problem.

Wndchrm is also sensitive to the observation angle of the galaxy, so if a certain face-on model is used, it cannot be expected that Wndchrm will classify edge-on galaxies that match the model. In that case, edge-on models should be used to train Wndchrm. 

\section{Results}

As described in Section~\ref{data}, fifty four target images of galaxy mergers imaged by SDSS were used in this study, and 230 simulated mergers were generated by SPAM as described in Section~\ref{simulation} for each of the target SDSS images. The 230 simulated galaxy mergers of each target were separated into training and test sets such that 210 simulated images of each target were used for training, and the remaining 20 were used for testing as described in Section~\ref{pattern_recognition}. The experiment was repeated 10 times such that in each run the simulated images were randomly allocated to training and test sets. The classification accuracy was measured by the percentage of simulated test galaxy merger images that were associated by the algorithm with the class of simulated images generated from the same target merger image. Results show that 51\% of the test simulated galaxy merger images were classified correctly, which is significantly higher than $\sim$1.9\% of random classification.

As discussed in Section~\ref{pattern_recognition}, the main purpose of the study is not necessarily to classify galaxies, but to quantify and measure the morphological similarities between systems of interacting galaxies. However, the fact that the simulated galaxy merger images can be classified with accuracy far higher than random shows that the morphological analysis used in the study is informative, and can be used to measure visual similarities between different forms of interacting galaxies despite the complex morphology of the structures.

\begin{figure*}[hp]
\centering
\includegraphics[scale=1.00]{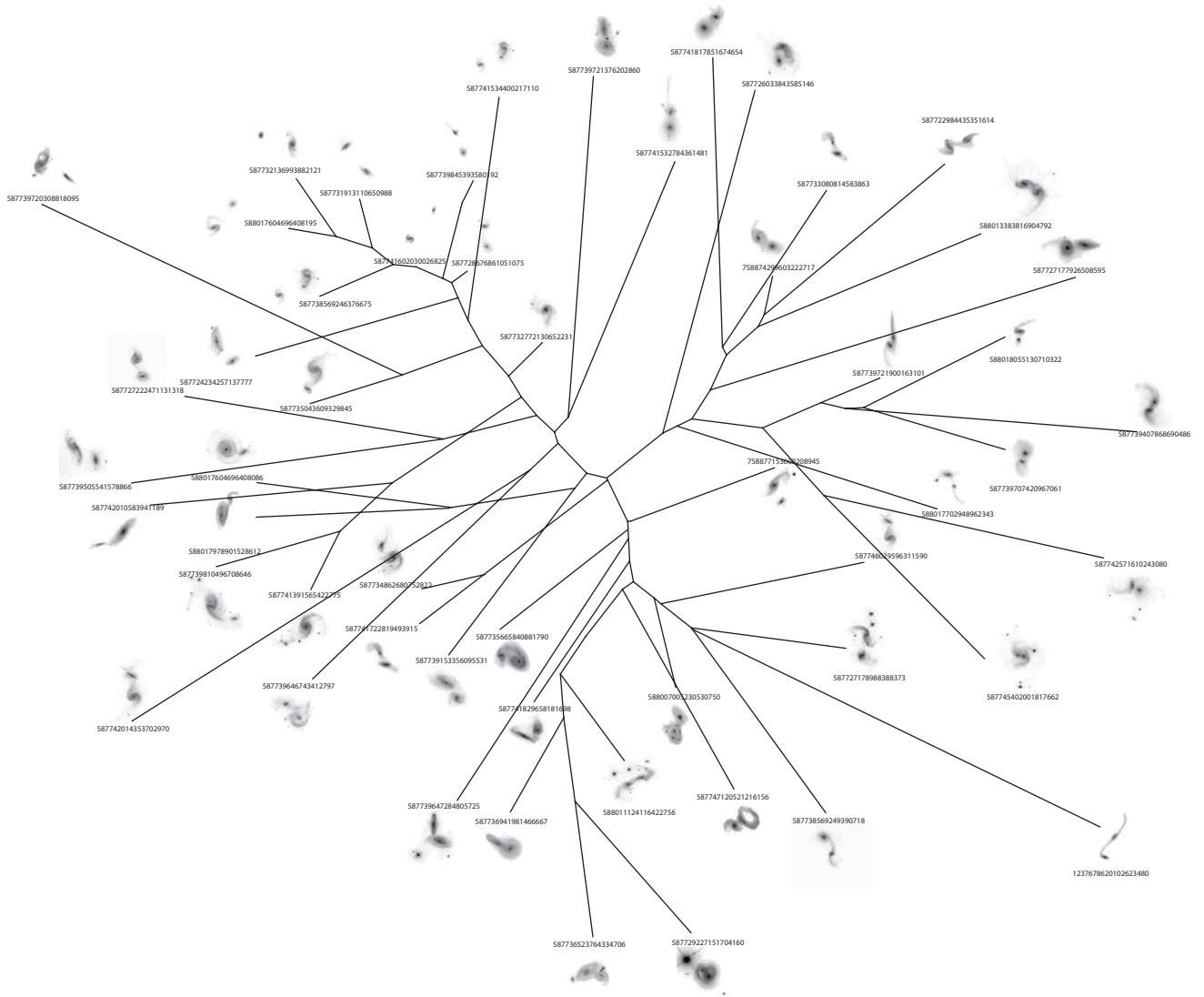}
\caption{Computer-generated phylogeny of the galaxy mergers, created using the simulated models generated by SPAM for each of the galaxy merger images in the graph. The numbers are the SDSS object IDs of the galaxies.}
\label{tree}
\end{figure*}

To measure and visualize the similarities between the images of interacting galaxies, the similarities were measured between the 54 classes of simulated merger images as described in Section~\ref{pattern_recognition}, such that each class contains the 230 simulated images. Figure~\ref{tree} shows the similarities between the target SDSS merger images, such that each target image is analyzed using the 230 simulated images that were generated for it by SPAM as described in Section~\ref{simulation}.

\begin{figure*}[hp]
\centering
\includegraphics[scale=1.4]{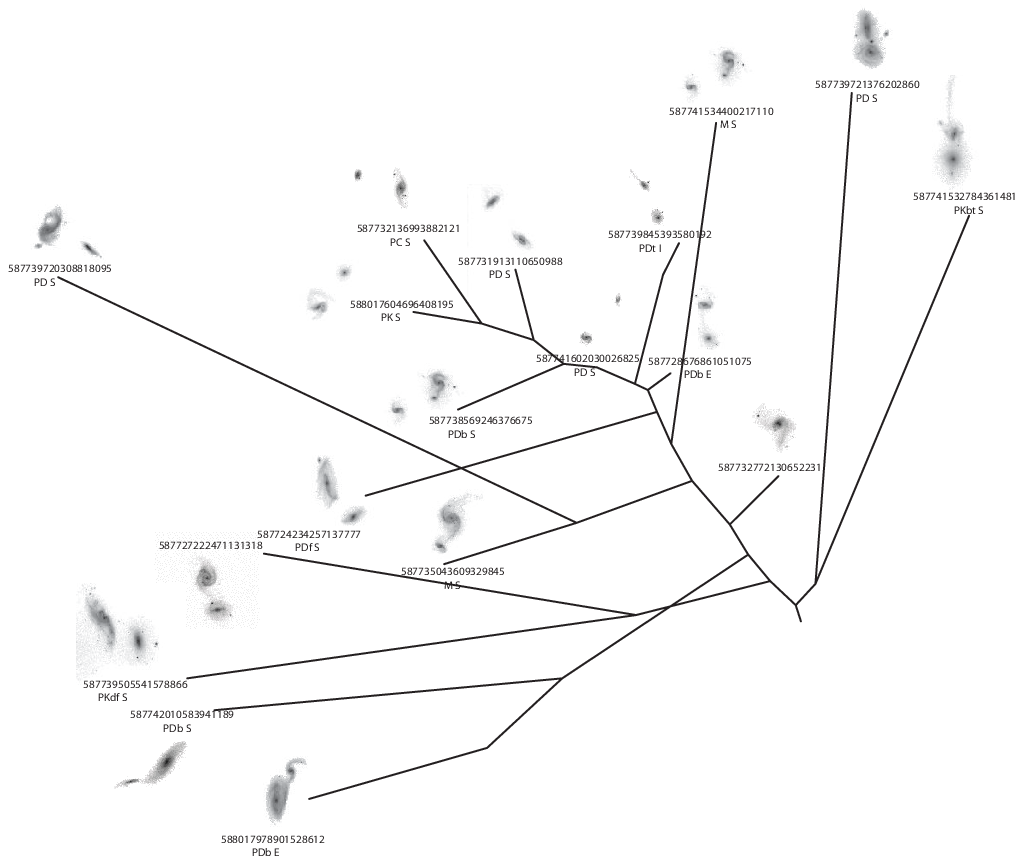}
\caption{The top left part of the similarity tree of Figure~\ref{tree}}
\label{top_left}
\end{figure*}

As the graph shows, the galaxies are grouped by their morphology. For instance, all galaxies at the upper left part of the graph are interactions between two spiral galaxies such that one is larger than the other. Figure~\ref{top_left} shows a closer view of that part of the tree, and also displays the VV catalog notation of their morphological classification. The center of the tree contains several spiral galaxies of about the same size and distance from each other. The right part of it contains two galaxies that are closer to each other compared to the other galaxies, and its left part is populated mostly with mergers such that one spiral galaxy is larger than the other. One of them is also galaxy 587739720308818095, which was positioned somewhat in separation from the other galaxy mergers. The positions of the mergers on the tree also feature approximate grouping by their VV catalog classification, showing similarity to the human crafted VV catalog.

\begin{figure}[hp]
\centering
\includegraphics[scale=1.1]{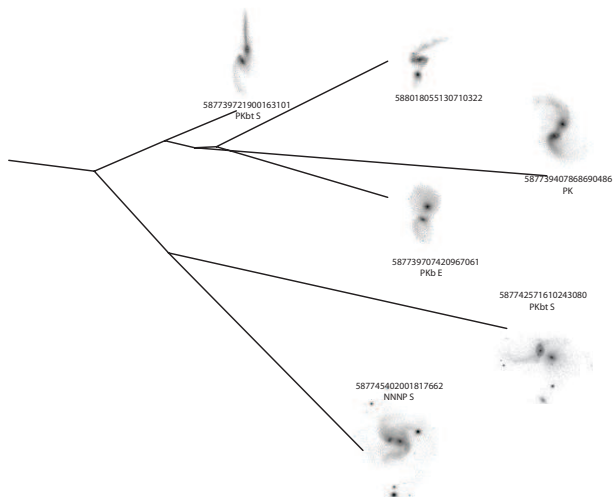}
\caption{The right part of the similarity tree of Figure~\ref{tree}}
\label{center_right}
\end{figure}

Figure~\ref{center_right} shows the right part of the phylogeny of Figure~\ref{tree}. In that part of the tree the two merging galaxies are also spiral, but closer to each other and at least one of the galaxies in each merger have long arms. Differences can also be noticeable between the mergers of the upper branch and the galaxy mergers on the lower branch of the tree of Figure~\ref{center_right}. The mergers 587742571610243080 and 587745402001817662 are larger in size, and each have one longer arm. Like in Figure~\ref{top_left}, the VV catalog classifications of the mergers in the graph is also consistent with the VV catalog.

\begin{figure}[hp]
\centering
\includegraphics[scale=0.95]{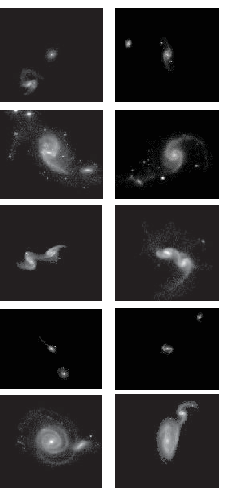}
\caption{Examples of pairs of galaxies placed close to each other by the algorithm}
\label{neighbors}
\end{figure}

The analysis can also be used to identify similar pairs of of mergers inside the dataset. For example, Figure~\ref{neighbors} shows examples of galaxies that were placed by the algorithm close to each other. As the figure shows, there galaxy mergers are also visually similar to each other, showing that the morphology measure used in this study can be used for automatically associating mergers of similar morphology in datasets of galaxies.

One of the downsides of the proposed method is its computational complexity. While generating the simulated galaxy mergers is a quick process, the bottleneck of the analysis is computing the WND-CHARM numerical image content descriptors for each simulated galaxy, which takes $\sim$60 seconds using a 2.4GHz AMD Opteron processor for each 256$\times$256 simulated image \citep{Sha08a}. Therefore, since each target galaxy merger image is represented by 230 simulated images, each target galaxy requires almost four hours of processing of a single core, and all 54 target galaxies can be processed within $\sim$8.6 days. Once the system is trained, each new galaxy is tested using 20 images, and therefore a new galaxy will take $\sim$20 minutes to compute using a single core. However, since the extraction of numerical image content descriptors can be easily parallelized with little overhead \citep{Sha08a}, systems with multiple processors and multiple cores can be used to significantly reduce the response time. In this study the galaxies were processed using 160 cores.

When processing a new galaxy takes $\sim$20 minutes using a single core, a mid-size computing cluster of 1000 can process 1M galaxies in about two weeks. Implementing the Wndchrm algorithm to run on a GPU by compiling it with CUDA or OpenCL will also dramatically improve the response time of the algorithm, and will allow it to process even larger datasets of galaxy images.

\section{Conclusion}

Digital sky surveys have been becoming increasingly important, reinforcing the need for computational tools that can automate the analysis of large databases of astronomical images. Galaxy morphology is important since the morphology of galaxies carries important information about the early, present, and future universe. Interacting galaxies feature complex morphology, and therefore automatic analysis of galaxy mergers is a challenging task.

In this paper we described an unsupervised machine learning method that can analyze interacting galaxies by their morphology. The method works by first generating a large number of simulated galaxies from each target galaxy merger image, and then applying complex morphological analysis. The source code is freely available at:\newline http://vfacstaff.ltu.edu/lshamir/downloads/ImageClassifier.

The problem of unsupervised analysis of galaxy morphology is a challenging problem in pattern recognition. This work demonstrates the usefulness of a similarity measure computed from image features. One application of the current work is to compute the features of new merger image and then determine its classification in the current phylogenic structure by finding the best similarity score. As larger populations of images of mergers become available, redoing the entire analysis, with corresponding simulations, will allow for creation of a new phylogeny. 

In its most simple form, the ability to automatically analyze images of galaxy mergers will allows generating large merger catalogs from databases such as LSST, where manual generation of such catalogs by inspection of each galaxy is highly impractical even with the use of the power citizen science. In its more advanced forms it can also associate newly imaged galaxies with their morphological structure as defined by existing schemes such as the Arp catalog \citep{Arp66}. This will provide more specific catalogs of galaxy mergers that go beyond merely the broad morphological types such as mergers or on-mergers, but will lead to a high-resolution database of morphology in which each galaxy merger is assigned with its similarity to defined structures, providing much larger set of samples for each type and allowing statistical analysis of the populations based on redshift, color, etc’.

Besides classifying images and simulations of multiple galaxies, the similarity measure could be the essential component of an automated fitness function for calculating how well a simulation matches an actual image of interacting galaxies. Such a fitness function could be combined with standard optimization techniques to automatically determine the best-fit simulation parameters for recreating the morphology of interacting galaxies. Future publications will include a study of how orbital dynamics is related to morphological classification based on this study and the results of the Merger Zoo project \citep{Hol13}.

\section{Acknowledgment}

The research was supported by NSF grant number 0941610. The SDSS is managed by the Astrophysical Research Consortium for the Participating Institutions. The Participating Institutions are the American Museum of Natural History, Astrophysical Institute Potsdam, University of Basel, University of Cambridge, Case Western Reserve University, University of Chicago, Drexel University, Fermilab, the Institute for Advanced Study, the Japan Participation Group, Johns Hopkins University, the Joint Institute for Nuclear Astrophysics, the Kavli Institute for Particle Astrophysics and Cosmology, the Korean Scientist Group, the Chinese Academy of Sciences (LAMOST), Los Alamos National Laboratory, the Max Planck Institute for Astronomy (MPIA), the Max Planck Institute for Astrophysics (MPA), New Mexico State University, Ohio State University, University of Pittsburgh, University of Portsmouth, Princeton University, the United States Naval Observatory and the University of Washington. We also wish to acknowledge the contributions of the Citizen Scientist volunteers at Galaxy Zoo Mergers for creating these models. A full list of the contributors fo this project can be seen at http://mergers.galaxyzoo.org/authors

\end{document}